\documentclass[aps,prl,twocolumn,groupedaddress]{revtex4-1}

\usepackage{graphicx}
\usepackage{dcolumn}
\usepackage{bm}

\begin{document}

\title{Interface effects in d-wave superconductor-ferromagnet
junctions\\
in the vicinity of domain walls}

\author{T. Kirzhner and G. Koren}

\affiliation{Physics Department, Technion-Israel Institute of Technology,
Haifa 32000, Israel}

\date{\today}

\begin{abstract}
Measurements of the differential conductance spectra of
$YBa_2Cu_3O_{7-\delta}-SrRuO_3$ and
$YBa_2Cu_3O_{7-\delta}-La_{0.67}Ca_{0.33}MnO_3$ ramp-type
junctions along the node and anti-node directions are reported.
Interpretation of the results in terms of crossed Andreev
reflection effect and induced triplet superconductivity are
discussed. The results are consistent with a crossed Andreev
reflection effect only in $YBa_2Cu_3O_{7-\delta}-SrRuO_3$
junctions where the domain wall width of $SrRuO_3$ is comparable
with the coherence length of $YBa_2Cu_3O_{7-\delta}$. No such
effect was observed in the
$YBa_2Cu_3O_{7-\delta}-La_{0.67}Ca_{0.33}MnO_3$ junctions, which
is in line with the much larger ($\times$10) domain wall width of
$La_{0.67}Ca_{0.33}MnO_3$. We also show that crossed Andreev
exists only in the anti-node direction. Furthermore, we find
evidence that crossed Andreev in $YBa_2Cu_3O_{7-\delta}$ junctions
is not sensitive to nm-scale interface defects, suggesting that
the length scale of the crossed Andreev effect is larger than the
coherence length, but still smaller than the
$La_{0.67}Ca_{0.33}MnO_3$'s domain wall width.\\

\end{abstract}

\maketitle

The proximity effect between a ferromagnet (F) and a
superconductor (S) has attracted much attention in the past few
years. Many scattering processes can occur at the S-F interface,
and a full understanding of these effects is still lacking. One
such process is the non-local crossed Andreev Reflection effect
(CARE), where an incident electron is reflected as a hole into a
spatially separated electrode while a cooper-pair is created in
the superconductor ~\cite{Byers}. In this process the electron and
hole remain in an entangled coherent state as long as they do not
scatter inelastically. Another possible process is the proximity
induced triplet superconductivity (PITS) in F in the vicinity of
inhomogeneities \cite{Bergeret,Volkov}. The CARE process can be
detected experimentally by measuring the conductance of spatially
separated normal metal (N) or ferromagnetic (F) electrodes coupled
to a superconductor, while the PITS process can be measured only
in S-F contacts. Previous studies show that CARE is possible only
when the spatial separation of the metallic electrodes is of the
order of the superconducting coherence length $\xi_S$
~\cite{Deutscher}. Both CARE and PITS can exist in the vicinity of
domain walls in S-F junctions. CARE seems to occur as long as the
domain wall (DW) width is comparable to $\xi_S$
~\cite{Melin,Aronov,Asulin}, while the existing theories for PITS
do not necessitate such a stringent requirement \cite{Volkov,KB}.

Since CARE is a spin dependant process, half metal ferromagnetic
leads with non-zero spin polarization can cause favoring of the
CARE process under certain conditions ~\cite{Beckmann}. In the
case of fully polarized ferromagnetic leads with anti-parallel
spin configuration, spin up electrons in one electrode can be
reflected as spin down holes in the other electrode, allowing
CARE, while in the case of parallel spin configuration, CARE
cannot exist. It was shown that the DW structure of a ferromagnet
can greatly modify the proximity effect of an S-F system
~\cite{Aronov,Asulin}. CARE effect is expected when the
ferromagnet DW width is of the same order of magnitude as $\xi_S$,
where non-local Andreev reflection is possible. When the
ferromagnet is not fully polarized, local Andreev reflection is
also possible and contributes to the junction conductance.\\

To this day, most of the research related to CARE and PITS was
focused on the case of s-wave superconductors. In the case of
d-wave superconductors, better understanding is still needed.
These cases were discussed theoretically by Herrera et al.
~\cite{Herrera} and by Volkov and Efetov \cite{VE}, but only a few
experimental studies had been reported ~\cite{Aronov,Asulin}. In
~\cite{Herrera}, several unique properties of CARE were
investigated in the configuration of two metal electrodes in
contact with a d-wave superconductor. Surprisingly, it was found
that CARE is a long range effect in d-wave superconductors and can
occur with electrode separation of up to $\sim 5\xi_S$, as opposed
to only $\sim \xi_S$ in s-wave superconductors. In addition, the
angular dependance of CARE was found to be strongest in the
anti-node direction.\\

In this study we investigate signatures of CARE and PITS effects
in d-wave superconductors by using high quality ferromagnet -
d-wave superconductor (F-S) ramp-type junctions with $a-b$ plane
coupling (see inset (b) of Fig. 1). We prepared various
configurations of ramp junctions containing optimally doped
$YBa_2Cu_3O_{7-\delta}$ (YBCO) as the d-wave superconductor and
two different types of ferromagnets: $SrRuO_3$ (SRO) and
$La_{0.67}Ca_{0.33}MnO_3$ (LCMO). The junctions were prepared
along two different angular orientations, in the node and
anti-node directions. SRO and LCMO have been chosen as the
ferromagnetic electrodes to investigate the DW width dependance of
the studied effects. The DW width in SRO is $\sim$2-3 nm
~\cite{Marshall,Feigenson} which is comparable to $\xi_S\sim$2-3
nm of YBCO. LCMO has a much larger DW width, in the range of
$20-40\,nm$ at the low temperature regime ~\cite{Lloyd,Ziese}.
Typical domain widths of SRO and LCMO are of about 200 and 300 nm,
respectively \cite{Marshall,Lloyd}. It was also found in these
studies that the DW orientations of c-axis grown SRO and LCMO
films on (100) $SrTiO_3$ (STO) are along the (110) and (100)
directions, respectively. If we assume similar DW orientations
when grown on twinned c-axis YBCO, this corresponds to the nodes
and antinodes directions. This leads to many domain wall crossings
of the interface in our $5\,\mu m$ wide junctions (about 18 and 15
crossings in the antinode junctions with SRO and LCMO,
respectively, with the corresponding values of 25 and 11 in the
node oriented junctions). We found that the differential
conductance spectra of these junctions provide evidence for the
long range nature of the CARE effect, and agree well with the
theoretical predictions of the angular dependance of this
effect~\cite{Herrera}. Within the framework of the existing
theories for PITS, no agreement with the detailed experimental
data was found.\\

\begin{figure}
\includegraphics[height=6.5cm]{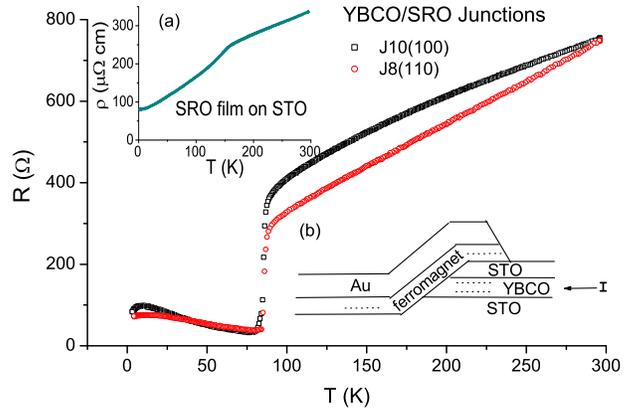}
\caption{\label{fig:srort}Resistance versus temperature of
YBCO-SRO junctions along the node (squares) and anti node
(circles) orientations. Insets: (a) $\rho$ versus T of a bare SRO film
on (100) STO. (b) A schematic ramp junction cross section.}
\end{figure}

A schematic diagram of a typical ramp junction is shown in inset
(b) of Fig.~\ref{fig:srort}. These junctions were prepared by a
multi-step process. The films were prepared by laser ablation
deposition, patterning was done by a water-less resist and deep UV
photolithography and etching by Ar-ion milling. Anti-node (100)
and node (110) oriented junctions where obtained by depositing
c-axis YBCO films on (100) STO wafers of $10\times 10\, mm^2$
area, with their sides parallel to either the (100) or (110)
orientations, respectively. The detailed preparation process is
described elsewhere ~\cite{Nesher}, and here we shall only briefly
summarize the main steps. First, the base electrode was deposited
which include a 60 nm layer of STO on top of an 80 nm thick YBCO
film. This bilayer was patterned on half of the wafer into ten 15
$\mu m$ wide bridges with a ramp angle of $\sim 35^\circ$, and
connections to the contact pads. The cover electrode was deposited
next and included either SRO or LCMO layers of 20 nm thickness
capped with an 80 nm thick gold layer for the leads and contacts.
Then the final patterning produced ten separated junctions of 5
$\mu m$ width on the wafer with four contacts each. Overall, in
the present study 12 such wafers were prepared, with the results
being reproducible in at least 3 junctions of each orientation and
type.\\

\begin{figure}
\includegraphics[height=6.5cm]{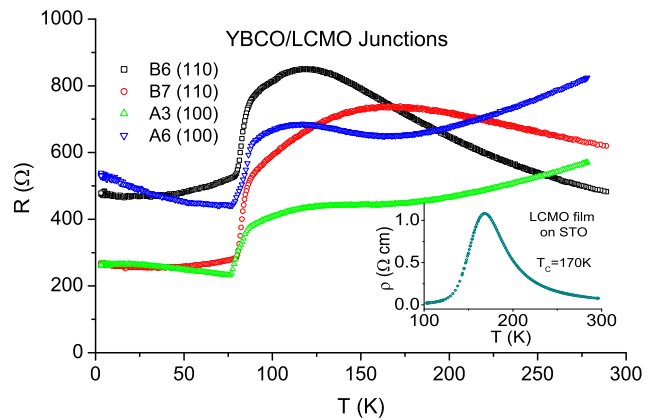}
\caption{\label{fig:lcmort}Resistance versus temperature of
YBCO-LCMO junctions along the node (squares and circles) and anti
node (up and down triangles) orientations. Inset: $\rho$ versus T of a
bare LCMO film on (100) STO.}
\end{figure}

Fig.~\ref{fig:srort} shows the R versus T curves of the YBCO-SRO
junctions. Inset (a) to this figure shows the $\rho$ versus T of a bare
SRO film on (100) STO where a clear signature of the ferromagnetic
transition at 150 K can be seen. It's hard to notice this
transition temperature in the main panel at around 150 K due to
the low SRO film resistance (20 nm thickness and $\rho\sim
0.5\,m\Omega$cm) compared to that of YBCO (80 nm thickness and
$\rho\sim 1.5\,m\Omega$cm). The low temperature resistance of
these junctions ($\sim$100 $\Omega$) is much higher than that of
the bare SRO films. The origin of this interface resistance
therefore, is apparently due to exposure of the ramps to ambient
air before the deposition of the cover electrode.
Fig.~\ref{fig:lcmort} shows typical R versus T curves of the
YBCO-LCMO junctions in the main panel together with the $\rho$ versus T
data of a bare LCMO film on (100) STO in the inset. The resistance
curves have a broad maximum at around 110-180\,K, which indicates
the transition temperature to ferromagnetism $T_c$ of the LCMO
layer. At low temperatures, the junctions have resistances of
200-600 $\Omega$ which are much higher than the typical LCMO film
resistance (see the inset of Fig.~\ref{fig:lcmort}). Similarly to
the SRO case, it is due to the formation of a more resistive layer
at the interface.\\

\begin{figure}
\includegraphics[height=6.5cm]{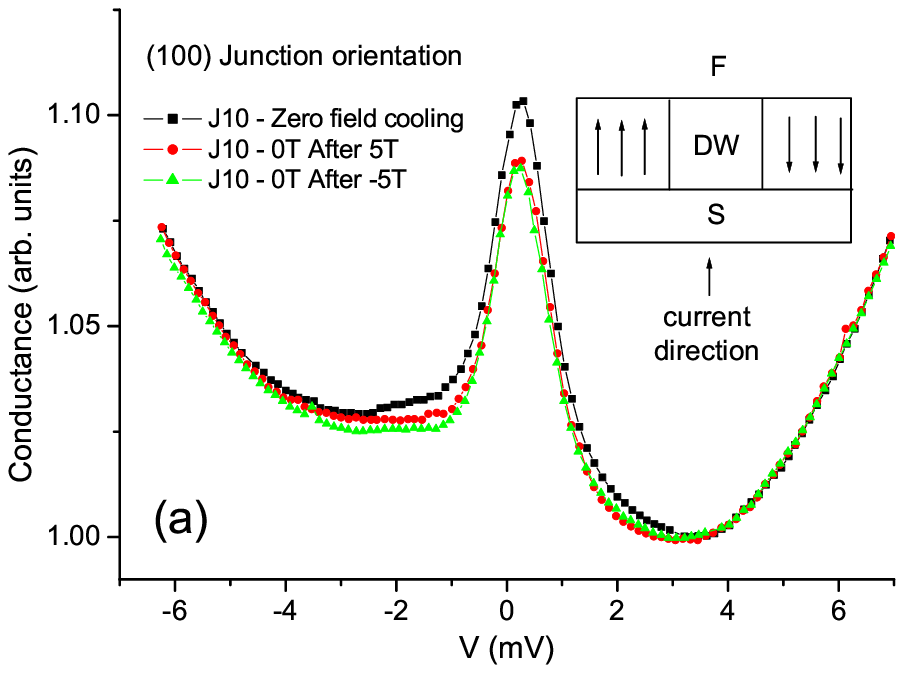}
\includegraphics[height=6.5cm]{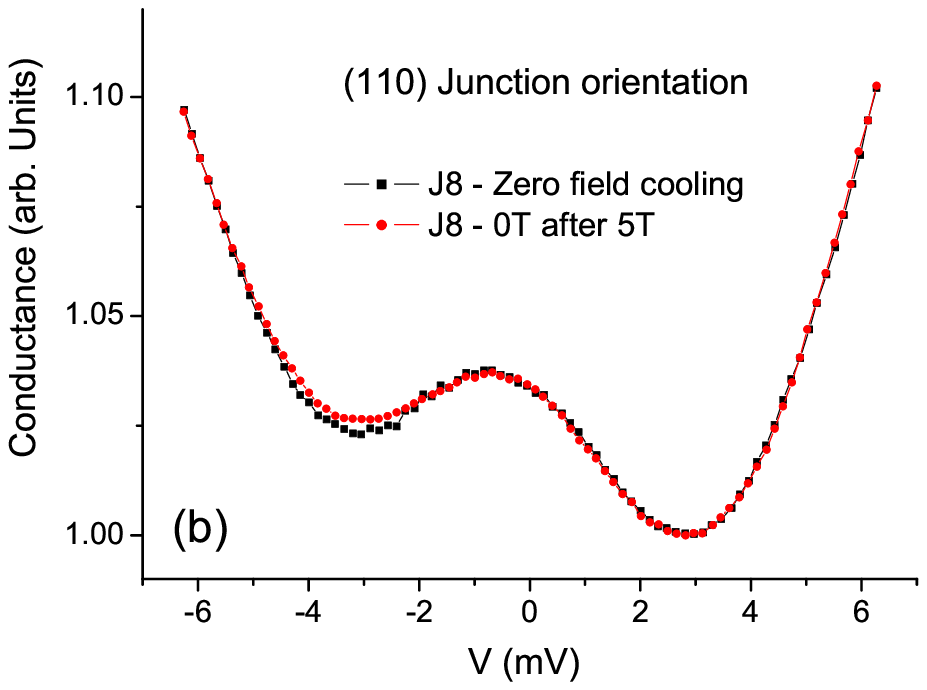}
\caption{\label{fig:sro}Conductance spectra at zero field of (a)
(100) and  (b) (110) oriented YBCO-SRO junctions. The squares
represent the zero field cooling spectra, while the circles and
triangles spectra show the results after field cycling to +5 T and
-5 T, respectively. Inset: The ferromagnet DW geometry in our S-F
junctions.}
\end{figure}

Fig.~\ref{fig:sro}(a) shows three normalized differential
conductance spectra of (100) oriented YBCO-SRO junctions at zero
magnetic field. One, after zero field cooling and the other two,
after applying fields of $\pm$5 T in a direction perpendicular to
the $CuO_2$ planes of YBCO. The conduction spectra show a clear
Zero Bias Conductance Peak (ZBCP) superposed on a parabolic background. As shown by Tanaka and
Kashiwaya for a d-wave superconductor ~\cite{Tanaka}, the ZBCP is
a result of Andreev bound states near zero bias which are formed
in the (110) (node) direction of the N-S junctions, while a
tunneling gap is excepted in the (100) (anti-node) direction. Many
theoretical and experimental results however
~\cite{Fogelstrom,Covington, Krupke, Cheng}, show that a ZBCP can
also be formed in the (100) direction, due to nm-scale interface
roughness, and that its strength can be comparable to that of a
ZBCP formed with a perfect (110) interface. Using AFM imaging of
the ramp (interface) of our junctions, we found that the surface
roughness is of about $\sim1-3$ nm, with no grain boundaries. This
indicates that the ZBCP we observed might originate in the node
bound states due to the interface roughness. Control experiments
made with YBCO-Au junctions oriented in the (100) and (110)
directions, showed ZBCP's with comparable strength in both
orientations, strengthening the assumption that nm-scale
roughness can lead to their formation.\\

As discussed before, the conductance of a superconductor coupled
to partially polarized ferromagnetic electrodes, has contributions
of both local and non local AR effects. SRO in the ferromagnetic
state has spin polarization of about 50\% ~\cite{Nadgorny}, thus
local AR is also possible. Since the domains width size ($\sim$200
nm) is so much larger than the DW width ($\sim$3 nm), the domain
walls configuration shouldn't affect the junction conductance
originated from local AR. To detect only the CARE effect, we
investigated how the domain wall structure affects the junction
conductance by application of various magnetic fields. Magnetic
fields of $\pm$5 T, which are much higher than the coercive fields
of both SRO (0.5 T) ~\cite{Klein} and LCMO (0.05 T) ~\cite{Ziese}
were applied, to make sure that the domain structure in the
ferromagnetic electrodes is changed. Hysteresis of the ZBCP height
can be seen in Fig.~\ref{fig:sro}(a) after application of these
magnetic fields and returning to zero field. This leads to a
reduction in the ZBCP height by $\sim$15\%, after this field
cycling process. A similar behavior was also seen previously by
Aronov and Koren ~\cite{Aronov} in (100) oriented S-F-S and S-F
junctions. To understand this behavior, we note that in zero field
cooling of the ferromagnet, many domains and domain-walls are
formed. Clearly, after the magnetic field cycling as described
above, the ferromagnet is much more oriented and the number of
domain-walls is significantly decreased. Considering the fact that
the SRO DW width is comparable to the superconducting coherence
length of YBCO, a large number of domain-walls crossing the
interface in S-F junctions should enhance the CARE conductance.
This agrees well with the present results of our
measurements.\\

It is well known that strong magnetic fields can also cause
suppression and/or splitting of the ZBCP ~\cite{Fogelstrom}. The
observed decrease in the junction conductance after field cycling
therefore, could be affected by the ferromagnet's remanent
magnetization field. This however is not the case, since the stray
magnetic field emanating from the SRO crystallites of $\sim$0.2 T
~\cite{Klein} is too weak to cause this effect. This was confirmed
in conductance measurement of the YBCO-Au junctions, where a very
little change of the ZBCP conductance was observed after the
application of a 0.2 T field.\\

Fig.~\ref{fig:sro}(b) shows conductance measurement results made
on a typical (110) oriented YBCO-SRO junction. No conductance
hysteresis after field cycling is found in this case, which is in
contrast to the results observed in the (100) junction of
Fig.~\ref{fig:sro}(a). According to Ref. ~\cite{Herrera}, in a
d-wave superconductor, CARE is strongest along the anti-node
directions where the order parameter amplitude reaches it's
maximum value. This prediction agrees well with our results. Our
results show that CARE which is a non local AR effect, has angular
dependance, in contrast to the local AR which is isotropic and
doesn't have angular variations. Considering the interface
roughness argument that was made above, one would expect that the
conductance increase due to CARE in the (100) and (110) oriented
junctions should be similar, as in the case of local AR. However,
since this conjecture is contrary to the present observations, it
is suggested that CARE is not affected by nm-scale roughness of
the junctions interface. This seems to indicate that long range
CARE correlations in YBCO which is a d-wave superconductor, are
responsible for the present results and are in line with the
theoretical predictions \cite{Herrera}.\\

\begin{figure}
\includegraphics[height=6.5cm]{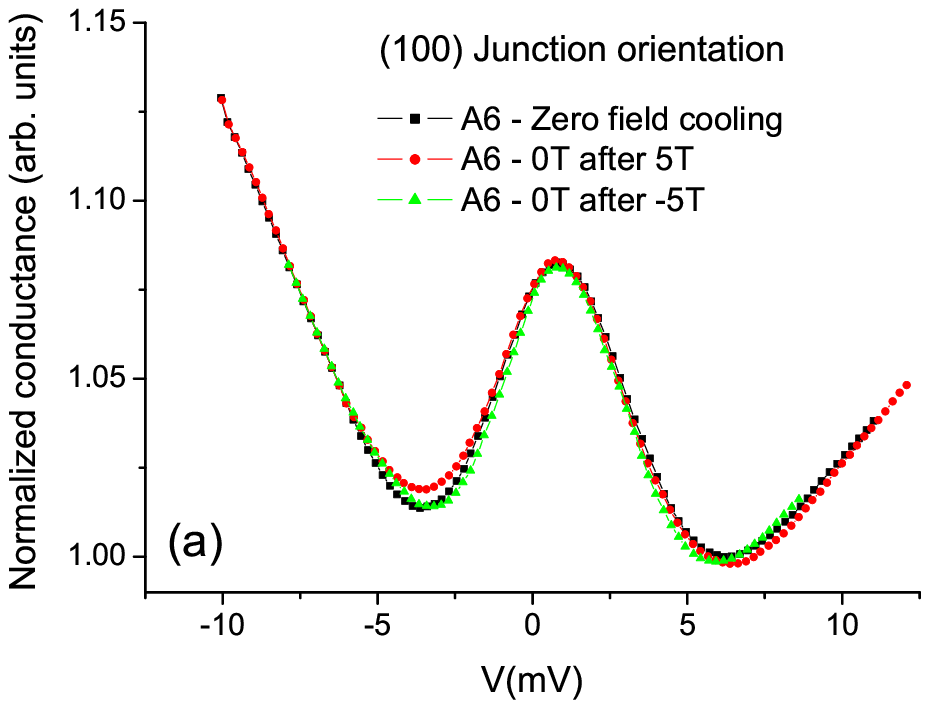}
\includegraphics[height=6.5cm]{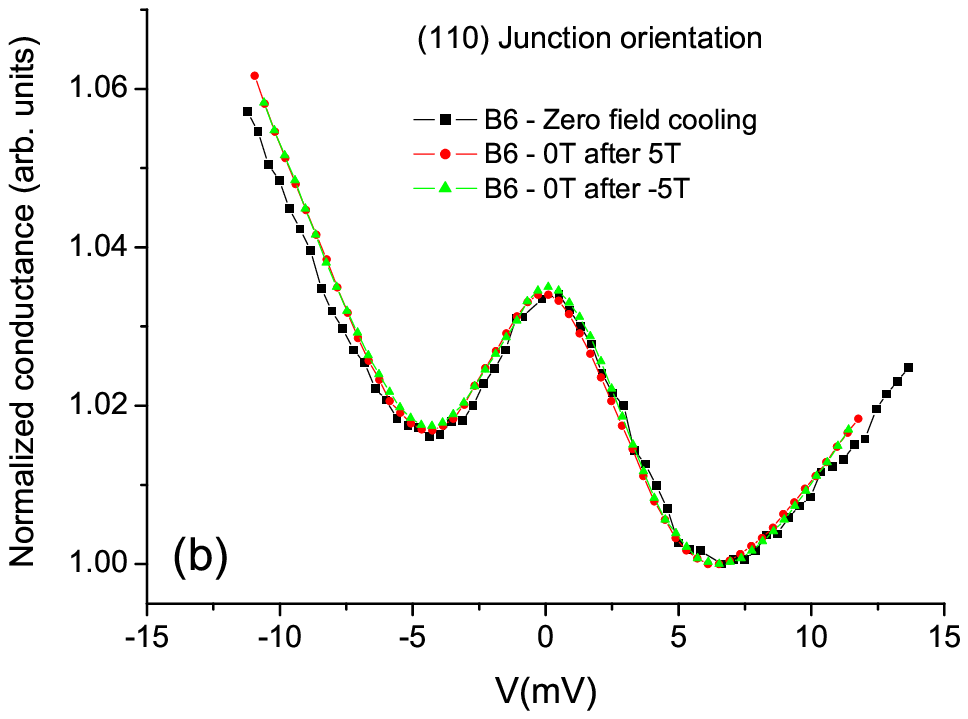}
\caption{\label{fig:lcmoc} Normalized zero field conductance
spectra of (a) (100) and  (b) (110) oriented YBCO-LCMO junctions,
under zero field cooling and after field cycling to $\pm$5 T and
back to zero field.}
\end{figure}

Fig.~\ref{fig:lcmoc} shows conductance spectra results taken in
(100) and (110) oriented YBCO-LCMO junctions. The conductance
curves had to be normalized because of the  high magneto
resistance of LCMO. The curves were normalized by division of the
data by the global minima value of each plot, to allow for a
convenient way to compare the ZBCP heights. Normalization by
subtraction of the global minima value of each plot was also tried
and yielded similar results (not shown). No hysteresis after field
cycling can be noticed in the ZBCP's height of any of the
YBCO-LCMO junctions. Considering that the LCMO DW width is much
larger than the coherence length of the superconductor $\xi_S$,
this result strengthens our argument that the hysteresis effect of
the ZBCP-height after field cycling seen in Fig.~\ref{fig:sro}(a)
is
due to CARE contribution rather than to other effects.\\

We shall now discuss our data in terms of PITS, the proximity
induced triplet superconductivity scenario. Bergeret et al. have
found that local inhomogeneities in the magnetization of a
ferromagnet in close proximity to a superconductor could induce a
triplet superconductive component inside the ferromagnet
~\cite{Bergeret}. This triplet component leads to a long range
proximity effect in F-S junctions and thereby increase their
conductance for any given barrier thickness. Local magnetization
inhomogeneities are naturally created in ferromagnets by the
formation of domains and domain walls. These domain walls could
theoretically induce triplet superconductivity in the ferromagnet
and therefore increase the junctions conductance. Our measurement
results could thus be interpreted in the framework of proximity
induced triplet superconductivity. In the (100) oriented YBCO-SRO
junctions of Fig.~\ref{fig:sro}(a) we found that the conductance
increases with the number of domain walls. A large number of
domain walls thus leads to many local inhomogeneities in the
magnetization, and this induces the triplet condensate component
in their vicinity. In our junctions geometry (see inset to
Fig.~\ref{fig:sro}(a)), theoretical calculations of a singlet
s-wave superconductor - ferromagnet junction show that the induced
triplet superconductive order in the ferromagnet persists even if
the DW width is much larger than the superconducting coherence
length \cite{Volkov,KB}. This prediction however, is in contrast
to our observations in the YBCO-LCMO junctions, where the number
of broad domain walls did not affect the junction conductance as
can be seen in Fig.~\ref{fig:lcmoc}. This indicates that either
the junction conductance in our case is not affected by triplet
superconductivity, or that it is affected by triplet
superconductivity but the DW width is an important factor in the
junction conductance. The latter case is supported by our
observations in YBCO-SRO junctions where the conductance is
enhanced with increasing number of narrow domain walls, as seen in
Fig.~\ref{fig:sro}(a).\\

Since in the present study a d-wave superconductor was used, it is
more appropriate to compare our results with the theoretical
treatment of Volkov and Efetov \cite{VE}. They considered an S-F
bilayer with S being a d-wave superconductor with the a-b planes
parallel to the interface, c-axis in the direction of the domain
walls in F and the current flowing along the c-axis direction. The
current in our junctions flows mostly along the a-b planes, but
due to the ramp angle a small current component also flows in the
c-axis direction and this allows a comparison with the results of
Ref. \cite{VE}. This paper predicts a maximal long range triplet
component if the DW planes lie in the antinode direction while no
triplet component exists if the DW planes lie in the node
direction. In the LCMO junctions the DW lie in the antinode
directions and thus a maximal triplet induced component is
expected. This should lead to a decrease in the low bias
conductance after magnetic field cycling due to the expected
decrease in the number of DW. Fig.~\ref{fig:lcmoc} shows that no
such effect occurs. In the SRO junctions the DW lie in the node
direction and the model predicts that no triplet component should
exist. Field cycling therefore should not affect the observed
conductance in this case, and this is contrary to our observations
(see Fig.~\ref{fig:sro}(a)). We thus conclude that in both type of
junctions, when there is a current component flowing in the c-axis
direction, the model results disagree with the experimental
results. This model however, might be more suitable to describe
the present observations if calculations were made with the
current flowing in the a-b plane. To summarize, while a proximity
induced odd frequency triplet s-wave component in the ferromagnets
can explain the enhancement of conductance at low bias (the ZBCP)
as seen in the experiments, the existing theories can not account
for the DW width dependence of the present results. The CARE based
interpretation however seems more suitable, since it agrees well
with all the current results and shows the correct dependence on
the DW widths.\\

In conclusion, we have found evidence for the existence of a CARE
effect in S-F junctions with the d-wave superconductor YBCO, but
proximity induced triplet superconductivity is also consistent
with some of our results. Only when the DW width was comparable to
the superconductor coherence length such as in the case of the
YBCO-SRO junctions, the low bias junction conductance was found to
be affected by the number of the domain walls in the ferromagnet.
In contrast, when the DW width was much larger than the
superconductor coherence length such as in the YBCO-LCMO
junctions, no effect on the junction conductance was observed.
These results can be interpreted as due to CARE, with a CARE
length scale on the order of the superconducting coherence length.
We observed that this effect is not affected by nm-scale interface
roughness, suggesting that it can also happen on length scales
larger than the superconductor coherence length. Moreover, we can
put an upper bound on this length scale which is found to be
smaller than the DW width of LCMO ($\sim 10\xi_S$). Finally, we
also show that the conductance change after magnetic field cycling
appears only in the (100) direction suggesting that
CARE in a d-wave superconductors is angle dependant.\\

{\em Acknowledgments:} We thank O. Millo for useful discussions.
This research was supported in part by the Israel Science
Foundation, the joint German-Israeli DIP project and the Karl
Stoll Chair in advanced materials.\\

\bibliography{AndDepBib.bib}

\end{document}